\documentclass[runningheads]{llncs}
\usepackage[T1]{fontenc}
\usepackage{xurl,hyperref}
\usepackage{amsmath,amssymb}
\usepackage[table]{xcolor}
\usepackage{multirow}
\usepackage{comment}
\usepackage{listings}
\usepackage{graphicx}
\usepackage{color}

\begin{document}
\title{Arbitrary Reduction of Validation Error for AI Decision Tests using Homomorphic AI and Repetition Codes\thanks{This paper is an extended and updated version of the talk presented at DSCNext Amsterdam 2026.}}

\author{Eric Filiol\inst{1}\orcidID{0000-0001-5101-8073} \and
	Jaagup Sepp\inst{2}}
\authorrunning{E. Filiol \& J. Sepp}
%
\institute{Independent Researcher - Retired Professor \\
	\email{efll@protonmail.com}\\
	\url{https://ericfiliol.site} \and
	Hope4Sec, Tallinn, Estonia\\
	\email{jaagup.sepp@hope4sec.eu}}
\maketitle             
\begin{abstract}
This paper presents new results and breakthrough obtained with the HbHAI techniques (Hash-based Homomorphic Artificial Intelligence) proposed in \cite{filiol0,sepp}.
HbHAI is based on a novel class of key-dependent hash functions that naturally preserve most similarity properties, most AI algorithms rely on. 
It enables to analyse and process data in its cryptographically secure form while using existing native AI algorithms without modification, with unprecedented performances compared to existing homomorphic encryption schemes and most notably compared to the same processing on corresponding plaintext data.

Two major results have been obtained further. First we enable to reduce the compression rate up to a factor of 10 thus allowing to process massive datasets while reducing the computation time and the energy footprint in the same order. Second, we show how it is possible to arbitrarily reduce the final validation error of AI-based decision tests by using repetition error-correcting codes. 

\end{abstract}
\section{Introduction}
In this paper, the “AI” term is used to describe all data analysis techniques (machine learning, deep learning, big data) to the exclusion of LLM (generative AI). For sake of concision, Homomorphic AI must be understood as Homomorphic Encryption for AI. 

As far as AI is concerned, most approaches require to use third-party environments such as clouds which provide suitable tools. The only other possibility is to use ``on-premises'' environments with suitable and skilled enough tech teams to deliver, run, maintain and to operate them. With cloud solutions the main security drawback lies on the fact that data owners do no longer control the access to data. 
Outsourcing data for the purposes to use dedicated AI tools as a service thus represents either a weakness or a critical risk. Indeed, we observe that a data has three essential ``vocations'':
\begin{itemize}
	\item to grow indefinitely (cost issues in terms of storage, computing time, bandwidth consumption),
	\item to be shared or accessed (and thus lead to misuse) with dubious third-parties (e.g. data brokers, national police or intelligence agencies),
	\item and, worse, to leak in the wake of attacks (for 2025, for instance refer to~\cite{mvizard}). It is worth noticing that this risk equally exists for ``on-premises'' environments.
\end{itemize}
The most effective protection is to be able to process data directly in encrypted form without using data under their plaintext form. In this way, in storage or during processing, any attacker or unauthorised third party will only have access to data in a form that cannot be exploited by them. This protection is called \textit{Homomorphic Artificial Intelligence} (HAI) \cite{he4ds} coming from the original research area of \textit{Homomorphic Encryption} (HE).

Another problem to address when processing data is the risk of error inherent in any AI test (which is, in fact, a standard statistical test) or validation error. There are two types of such errors: false positives and false negatives \cite{neyman-pearson}. These decision errors, even small, may have dramatic consequences in critical use-cases (medicine, defence, security\ldots). It is therefore a critical problem to minimize these errors as much as possible. At the present time, approaches to reduce errors imply stronger preprocessing of data, collecting more data of better quality thus implying higher computing power. But the natural variability of a given population from sample to sample makes this issue quite impossible to solve efficiently. 

At the end of 2020, we launched a collaborative project to develop a totally new and disruptive approach to homomorphic encryption applied to AI. The aim was to start from scratch and design a homomorphic data analysis scheme, called HbHAI (standing for \textit{Hash-based Homomorphic Artificial Intelligence}) that would provide at least the same level of cryptographic security for the data, while removing the constraints and limitations in existing HE schemes.   

For the time being, the HbHAI scheme is not public because it is not yet protected in terms of intellectual property. Moreover the industrial exploitation is still pending. Datasets have been made public to the community \cite{sepp} in order to make possible an external analysis, and other will be as soon as possible. 

In this paper, we provide new results that confirm the very significant potential of HbHAI techniques:
\begin{itemize}
	\item we achieve a significant reduction of data and models size and hence of computing time and energy consumption, up to a factor of 10. This enables homomorphic processing of massive encrypted datasets using AI (comprising millions of individuals, each described by hundreds of thousands of features) with classical algorithms/tools; 
	\item thanks to the very nature of HbHAI, we have been able to arbitrarily reduce the validation error for any AI test, by means of repetition error-correcting codes.
\end{itemize}
The paper is organised as follows. In Section~\ref{sec1}, we summarize the main features of HbHAI techniques and the results obtained so far on a few datasets. In Section~\ref{sec2}, we present new results on HbHAI-protected datasets. Section~\ref{sec3} explains how we succeeded to reduce the validation error of any AI test. Finally Section~\ref{sec4} summarizes our results and mentions the future works and development for HbHAI.
\section{Homomorphic AI Hash Functions (HbHAI)} \label{sec1}
\subsection{Formal Definition and Features}
In order for the paper to self-contained, this section summarizes the main definitions and features of HbHAI techniques. The formalization has been published in \cite{filiol0}. First datasets and use-cases have been presented in \cite{sepp}. A first technical evaluation on those datasets has been presented at CyberWiseCon 2025 \cite{filiol2025}.

In order to provide cryptographic primitives suitable for homomorphic AI, a new class of keyed-hash functions has been designed.. The use of keyed hash functions aims at the same to provide a strong cryptographic security and a significant data size and computing time reduction for existing AI algorithms generally used.
 
 \begin{definition} (HAI Hash Function Class) \cite{filiol0} \label{defhai}
A keyed hash function for HAI applications is a function $H_{K, \delta}$ parametrized by a secret key $K$ and a compression rate $\delta$, which has, as a minimum, the following two properties:
	\begin{enumerate}
		\item \textbf{Compression} — $H_{K, \delta}$ maps an input $x$ of arbitrary finite bit length $n$, to an output $H_{K, \delta}(x)$ of bit length $\frac{n}{\delta}$.
		\item \textbf{Ease of Computation} — Given $H_{K, \delta}$ and an input $x$, $H_{K, \delta}(x)$ is easy to compute.
		\item \textbf{Similarity Preserving} - For a given similarity measure $S$ and any three objects $x, x', x''$ then we have, 
		\[S(x, x'') < S(x, x') \Leftrightarrow S(H_{K, \delta}(x), H_{K, \delta}(x'')) < S(H_{K, \delta}(x), H_{K, \delta}(x'))\]. 
	\end{enumerate}
\end{definition}
This definition considers similarity instead of the more restricting concept of distance. Most AI techniques, not to say all, are based in a way or another on the central concept of similarity (between objects). Most similarity measure can be converted to distance but not all (for instance Cosine similarity). 

\begin{figure}[h]\label{fig0}
	\begin{center} 
		\includegraphics[width=10.3cm]{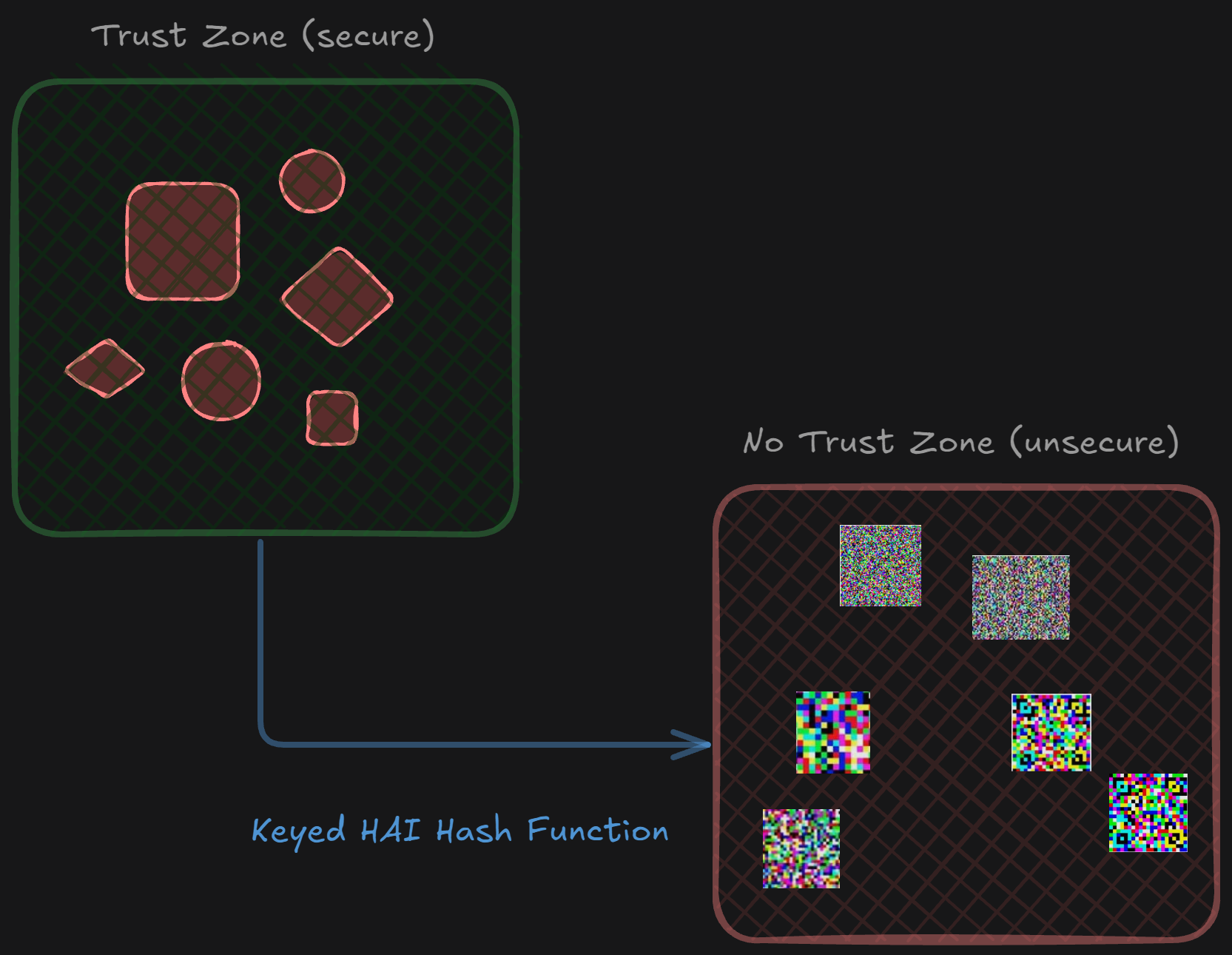} 
	\end{center}
	\caption{Illustrative description of HbHAI principle} 
\end{figure}

In order to illustrate this more intuitively, we consider Figure~\ref{fig0}.
Once cryptographically protected, it is no longer possible to guess which form is a square or a circle (their plaintext description) but similarities between plaintext forms are preserved beyond cryptography. 

The interested reader can refer to \cite{filiol0} for a formalization of threat model (with respect to AI) we must consider and which security properties must be fulfilled. 

The use of hash functions (non-injective transformations) invalidates the concept of decryption in HE. With HbHAI, a different approach has been defined. Only the owner of the data knows the correspondence between the unencrypted and encrypted versions of an object/individual in a dataset, thanks to the index of each individual (after decrypting the index eventually).
It is possible \cite{filiol0} to transpose the results of AI algorithms obtained on the encrypted version of a dataset to the unencrypted version of the dataset, using only the indexes (clustering, classification\ldots). The decryption operation is therefore no longer necessary, in the specific context of AI.

In addition to strong security requirements, HbHAI specifications also include other requirements, the most essential of which are as follows: 
\begin{description}
	\item[Frugality] \textit{i.e.} minimizing the computational resources required both to calculate the models and to operate them. This concerns both a significant reduction in data size (parameter $\delta$) and a significant reduction in computation time. The main benefit is reduced energy and ecological footprints and an enhanced operability in critical and constrained systems (drones, embedded systems, IoT).
	\item[Portability] Data protected by HbHAI must to be processed by existing algorithms, in their original form, without rewriting (Keras, Tensorflow, custom implementation of classic AI algorithms\ldots).
	\item[Sovereignty and Independence] HbHAI technology must be operated in constrained environments and not on cutting-edge technologies (subject to US embargoes and export controls). In this respect, all our implementations and experiments have been performed on an Odroid H4 Ultra single board computer with 8-core CPU (architecture Alder Lake N, 32 Gb DDR5-4800 \& 1Tb SSD), running Linux Pop!\_OS or Linux OpenSuse 15.4 (as a development environment with gcc compiler and GMP library \cite{gmp}).
\end{description}   
\subsection{Current Performance Summary}
Two datasets have been provided with the following parameters: a 256-bit secret key $K$ and $\delta \in  [3, 6] \subset \mathbb{R}$. These two datasets are presented in \cite{sepp} and are now available on Hope4Sec's website. They have been specifically designed to evaluate most AI algorithms and problems.
\begin{table}
	\centering
	\begin{tabular}{|c|c|c|} \hline \rowcolor{lightgray} 
		Features                                           &       Dataset 1          &           Dataset 2           \\ \hline \hline
		Data type                                          &  \; \;Cyber data \; \;   &   \; \;Greyscale Images\; \;  \\ \hline
		Number of objects (training)                       &        2,000             &            60,000             \\ \hline
		Number of objects (validation)                     &          200             &            10,000             \\ \hline
		Number of features                                 &        49,955            &              N/A              \\ \hline
		Number of clusters/classes                         &           2              &              10               \\ \hline
		\; \;Original dataset size (Tr.+Val.)\; \;         &         14 Mb            &           30.3 Mb             \\ \hline \hline
		\; \;HbHAI-protected size (Tr.+Val.)\; \;          & \multirow{2}*{4.70 Mb}   &      \multirow{2}*{11.3 Mb}   \\ 
		($\delta = 3$)                         &                          &                               \\ \hline
		\; Computing time reduction \; \;      & \multirow{2}*{2.99}      &      \multirow{2}*{0.2}   \\ 
		($\delta = 3$)                         &     &                               \\ \hline
		\; Validation Performance \; \;        & \multirow{2}*{No loss}    &      \multirow{2}*{0.9534}   \\ 
		($\delta = 3$)                         &     &                               \\ \hline 
		Algorithms used                        &  Clustering and k-NN     &             Random forest (sk.learn)              \\ \hline \hline
		\; \;HbHAI-protected dataset size (Tr.+Val.)\; \;  & \multirow{2}*{2,34 Mb}   &     \multirow{2}*{5.2}        \\  
		($\delta = 6$)  &                          &                               \\ \hline
		\; Computing time reduction \; \;      & \multirow{2}*{5.98}      &      \multirow{2}*{0.2}   \\ 
		($\delta = 6$)                         &     &                               \\ \hline
		\; Validation Performance \; \;        & \multirow{2}*{No loss}    &      \multirow{2}*{0.938}   \\ 
		($\delta = 6$)                         &     &                              \\ \hline                       
		Algorithms used                        &  Clustering and k-NN     &             Random forest (sk.learn)              \\ \hline
	\end{tabular}
	\medskip
	\caption{Features of datasets used and performance results. ``No loss'' means that results are identical on HbHAI-protected and plaintext datasets.}	\label{tab1}
\end{table}
\begin{description}
\item[Dataset 1] This dataset gathers non public data coming from the cybersecurity domain. This two-class dataset is intended to test unsupervised learning (clustering) but also classification (identifying to which class new objects belong). Another aim is to evaluate the performances of HbHAI on massive datasets. in this respect, each individual (object) is described by 49,955 different features (categorical features). It includes one training set containing 2,000 files/individuals and one validation set containing 200 individuals. 
\item[Dataset 2] The second dataset is the Fashion-MNIST~\cite{dataset2} which is a dataset of Zalando's article images, consisting of a training set of 60,000 examples and a test set of 10,000 examples. Each example is a 28x28 greyscale image, associated with a label from 10 classes. Fashion-MNIST 
dataset's purposes is for benchmarking machine learning and deep learning algorithms. 
\end{description}
Table~\ref{tab1} summarizes the main features of those two datasets~\cite{sepp} as well as the performance results we have obtained so far~\cite{filiol2025}.

\textbf{Performance and computing time results are given in comparison to the same processing on the corresponding plaintext datasets.}

We have observed that HbHAI indeed provides an efficient and real preservation of AI algorithms efficiency/accuracy on HbHAI-protected data. We confirm that no information loss has occurred and that validation accuracy is very well preserved.
We also confirmed that HbHAI technique really enables to work with "off-the-shelf" software/tools/libraries without modification (up to parameters). But the clear potential and power of HbHAI can really be exploited with a dedicated, optimized implementation of classic AI algorithms 

In terms of computing performances and data size reduction, we confirm that the speed-up is compliant with data size reduction with dedicated, optimized classical AI algorithms. However with tools like scikit.learn or TensorFlow (the large overhead is due to some sort of internal inertia) only 20~\% computing time reduction has been measured. However is remains a huge speed-up compared to classic FHE versions of AI algorithms (one-million times slower compared to processing time on plaintext data \cite{jkun}).
\section{New Results on HbHAI-protected Datasets} \label{sec2}
We have developed HbHAI further especially for higher rate compression. We essentially worked on Dataset 1. the main reason lies in the fact that Dataset 2 contains individuals of rather small size and considering further compression would not make sense.

We now are able to work with compression rate $\delta \in  [3, 10] \subset \mathbb{R}$. Applied on Dataset 1, we still obtain the same results compared to processing on the plaintext version of Dataset 1. This confirms the potential of HbHAI techniques when working with massive datasets. We have implemented the same classical clustering clustering algorithms as in \cite{filiol2025}, in C with GMP library.

For $\delta = 10$, the final datasize of the dataset is 1.3 Mb while the computing time has been effectively reduced by a factor of 9.97.

This dataset is available on the Hope4Sec's webpage. We expect to find collaboration to apply and test HbHAI techniques on third-party datasets.
\section{Arbitrary Reduction of Validation Error} \label{sec3}
\subsection{Repetition Error-correcting Codes}
In order to solve this issue, we have borrowed concepts from the Error-correcting codes theory \cite{ecc-th}. We considered a particular class of linear codes called \textit{Repetition Codes} \cite[Vol. 1, p. 16]{ecc-hdbk}. Error-correcting codes aim at introducing redundant information to manage the noise during a communication. Each piece of information $u$ is encoded as a codeword $w_u$. During the transmission the noise transforms $w_u$ into $\widehat{w_u}$ which is then decoded as $w_u$ to retrieve the emitted information $u$.

Let us consider a $[n, 1, n]$ repetition code with $n = 2r + 1$ (code length is $n$, the corresponding linear subspace has dimension 1, the minimal distance is $n$ which directly determines the detection and correction capabilities). It works as follows.		
\begin{itemize} 
\item We consider 1-bit piece of information $u$ which is encoded into an $n$-bit string $w_u$ (repeated bits)
  \begin{itemize}
  	\item Let $n = 3$. We then encode $u = 0$ as $w_u = 000$ and $v = 1$ as $w_v = 111$. This code has only two possible codewords and hence is of dimension 1.
  \end{itemize}
\item In the most classical model of communication channel called the \textit{Binary Symmetric Channel}, each bit has a probability $p$ of being received incorrectly due to the noise effect.
  \begin{itemize}
  	\item For instance we suppose we receive (noisy) codewords $\widehat{w_u} = 0\textcolor{red}{1}0$ and $\widehat{w_v} = 0\textcolor{red}{11}$. Here bit in red are noisy (incorrect) bits.
  \end{itemize}
\item In order to decode received (error) word, we apply a majority decoding. We compute the Hamming weight $d_H(w)$ (which is the number of $1$ in the word $w$). If $d_H(w) < \frac{n - 1}{2}$ we decode $w$ as 0 otherwise 1. 
  \begin{itemize}
  	\item From the previous examples, then $\widehat{w_u}$ is decoded as $0$ (no residual error) and $\widehat{w_v}$ as 1 (one residual error since $v = 0$ is incorrectly decoded). 
  \end{itemize}
	\end{itemize}
While repetition codes are not the most economic ones (in terms of bandwidth, since a single bit is encoded as $n$-bit string), they however have strong correcting properties. To summarize, a $[n, 1, n]$ repetition code with $n = 2r + 1$
\begin{itemize}
	\item detects up to $(n - 1)$ errors,
	\item corrects up to $\frac{n - 1}{2}$ error and up to $n - 1$ erasure errors,
	\item and the residual error probability is  
	\begin{equation} \label{eq_proba}
		p_{\mbox{\tiny res.}} = 1 - \sum_{k = 0}^{\frac{n - 1}{2}} \binom{n}{k}p^k(1 - p)^{n - k}
	\end{equation} 
\end{itemize}
The last property is of high importance. It means that it is always possible to make the residual decoding error tends to 0 by simply increasing the codelength $n$ (under the assumption that the channel error probability $p$ remains the same). If $p < \frac{1}{2}$ then one can always increase the probability of success by increasing $n$.
\begin{example}
In the previous example, the noisy word $\widehat{w_v} = 0\textcolor{red}{11}$ has been wrongly decoded as $1$. So if we take $n = 5$, $w_v = 00000$. We suppose that the received word is $\widehat{w_v} = 0\textcolor{red}{11}00$ which is correctly decoded as $0$.
\end{example}
\subsection{Repetition Code and AI Decision Test with HbHAI}
We formalize the result of any decision test as a bit with $p$ equivalent to a decision error. Here $p$ represent somehow the ``noise'' introduced by a bad decision. Without loss of generality, we consider here a two-class decision tests (simple statistical test).
\begin{itemize}
\item Using HbHAI, we consider $n$ secret key $K_i$.
\item On the \textbf{same dataset} $\mathcal{D}$, we define $n$ decision tests $\mathcal{T}_i = H_{K_i, \delta}(\mathcal{D})$.
\item Without loss of generality, we also assume that tests $\mathcal{T}_i$ have the same error decision error $p$ and can be considered as independent variables.
\item We then apply the majority decoding on the $n$ test results to reduce the error.
	\begin{itemize} 
		\item With $p = 0.05$ and $n = 3$ we have $p_{\mbox{\tiny res.}} = 0.007$
		\item With $p = 0.05$ and $n = 5$ we have $p_{\mbox{\tiny res.}} = 0.0011$
	\end{itemize}
\end{itemize}	
We have successfully tested this approach on various cases (non public presently). On Dataset 1, we have reached a validation error less than $10^{3}$ for all compression rates for the classification part (using $k$-NN algorithm). 

In Equation~\ref{eq_proba}, we have supposed that $p$ remains constant over the different tests which moreover are 2-class decision tests. In fact, this equation can easily be generalised for different probabilities $(p_1, p_2,\ldots, p_n)$ and any number of decision classes.  

A number of datasets is about to be available on Hope4Sec's website among which the Zalando Fashion-MNIST dataset for $\delta = 6$ and three secret keys. The file format follows the original FASHION MNIST dataset \cite{deepwiki}.
\section{Conclusion \& Future Works} \label{sec4} 
In this paper we have presented further development and results for HbHAI, a new homomorphic encryption technique dedicated to AI. For massive datasets, we can now achieve a compression rate àf 10. By considering different secret keys, we also succeeded in reducing the validation error arbitrarily.

The further developments for HbHAI consider the following aspects and issues: 
\begin{itemize}
	\item  For massive datasets, we intend to increase the $\delta$ parameter to 20 (target: summer 2026). We already have the mathematical tools and we are now finalising the implementation and testing to confirm the theoretical results. 
	\item  We have noticed that results obtained on certain version HbHAI-protected Dataset 1 versions were better that those obtained on the plaintext version of the dataset. HbHAI techniques appear to have natural self-correcting properties, which we wish to confirm and analyse further.
	\item  We intend to generalize and explore the error reduction with other error-correcting codes, as well as for different models of communication channels. In particular, the Binary Symmetric Channel model may in some cases be not the most suitable one especially if there exist statistical dependencies do exist between decision tests.
\end{itemize}
We are interested in handling real-life cases and datasets provided by third parties. 

\begin{credits}
\subsubsection{\discintname}
There are no ethical issues. The authors do not have any competing interest of any kind. This research work was entirely self-financed. HbHAI techniques are the exclusive property of Hope4Sec.
\end{credits}

\begin{flushright}
	\textbf{S. D. G.}
\end{flushright}

\end{document}